\documentclass[twoside]{article}
\usepackage{fleqn,espcrc2}

\usepackage{graphicx}
\usepackage[figuresright]{rotating}

\newcommand{\AmS}{{\protect\the\textfont2
  A\kern-.1667em\lower.5ex\hbox{M}\kern-.125emS}}

\title{Wiedemann-Franz
violation in the vortex state of a $d$-wave superconductor}

\author{Wonkee Kim$^a$, 
        F. Marsiglio\address{Department of Physics, University of
        Alberta, Edmonton, Alberta, Canada T6G 2J1}%
        ~~and~~
        J. P. Carbotte\address{Department of Physics and Astronomy,
        McMaster University, Hamilton, Ontario, Canada, L8S~4M1}% 
}

\begin{document}

\begin{abstract}
We show that the Wiedemann-Franz law is violated in the
vortex state of a $d$-wave superconductor at zero temperature.
We use a semiclassical approach,
which includes
the Doppler shift on the quasiparticles
as well as the Andreev scattering from a random distribution
of vortices.
We also show that the vertex corrections to the electrical
conductivity due to the anisotropy of
impurity scattering become unimportant in the presence of
a sufficiently large magnetic field.
\vspace{1pc}
\end{abstract}

% typeset front matter (including abstract)
\maketitle

The Wiedemann-Franz (WF) law is one of the fundamental laws
which is well-obeyed in many solid state systems. The law is so universal
that its violations are considered to be very interesting phenomena.
The WF law
can be violated depending on the nature of impurity scattering and
the presence of a magnetic field in a $d$-wave superconductor.

Applying a magnetic field $(H)$ along the $c$ direction
to a $d$-wave superconductor, we study the in-plane
quasiparticle transport properties at 
zero temperature $(T=0)$. Since nodal quasiparticles
give the dominant contribution to the conductivity at $T=0$, we consider
only quasiparticles in the nodal areas in momentum space.
The magnitude of $H$ is assumed to be $H_{c1}<H\ll H_{c2}$, where
$H_{c1(2)}$ is the lower (upper) critical field.

At zero field $(H=0)$ and $T=0$, the D.C. conductivity
in a $d$-wave superconductor is universal \cite{lee,durst}; 
namely, it is independent of
the impurity scattering rate:
$\sigma_{00}=2(e^{2}/\pi^{2})(v_{f}/v_{g})$,
where $v_{f(g)}$ is the magnitude of the Fermi (gap) velocity.
Similarly, one can show that the thermal conductivity
is also universal:
$\kappa_{00}/T=(2/3)(v_{f}/v_{g}+v_{g}/v_{f})$
as $T\rightarrow0$.
Taking the ratio of $\kappa_{00}/T$ and $\sigma_{00}$, we immediately
find that the Lorenz number $L=\kappa/(T\sigma)\simeq L_{00}$,
where $L_{00}=\pi^{2}/(3e^{2})$, because $v_{f}\gg v_{g}$ for a
high-$T_{c}$ superconductor. This means that the WF law is obeyed.
%in a $d$-wave superconductor when $T=0$ and $H=0$. 
However,
it has been shown \cite{graf}
that the WF law is violated at finite $T$ 
and the violation can depend 
on the nature of the impurity scattering in the absence of a magnetic
field. In this work we show that the WF law is violated in the
vortex state at zero temperature based on the semiclassical approach,
which includes
the Doppler shift as well as Andreev reflection \cite{yu};
therefore, the total scattering is given by the impurity scattering and
the quasiparticle-vortex scattering \cite{kim}
which we take to be proportional to the magnetic energy $E_{H}$
with a proportionality constant $b$.

In Fig.~1, we plot the normalized Lorenz number $L/L_{00}$ as
a function of $E_{H}/\gamma_{00}$, where
\begin{equation}
E_{H}\simeq\sqrt{H/H_{c2}}~\Delta_{0}
\end{equation}
with a superconducting gap $\Delta_{0}$,
%$E_{H}$ is a magnetic field energy proportional to
%$\sqrt{H}$, 
and
$\gamma_{00}$ is the impurity
scattering rate in the absence of the magnetic field. If $L/L_{00}=1$,
the WF law is obeyed. Thus the difference between $L/L_{00}$ and unity
indicates the degree of the WF law violation. As one can see, the WF law
is violated for a sufficiently high field such that $E_{H}\gg\gamma_{00}$
regardless of the nature of the impurity scattering.
Note that in this plot (as well as in everything we have said so far),
we have considered only the bare bubble diagram 
for transport coefficients without
including the vertex corrections due to the anisotropy of the
impurity scattering. The effect of the vertex corrections
is the issue we investigate next. 
To do this we take into account
the ladder diagrams 
as was done in Ref.\cite{durst}.

We consider both the vertex corrections to the D.C.
conductivity and to the thermal conductivity.
As in Ref.\cite{durst}, we introduce the scattering anisotropy parameters
$R_{2}$ and $R_{3}$ for the adjacent-node and opposite-node scattering,
respectively. $R_{2}=R_{3}=1$ corresponds to the isotropic
case for which the vertex corrections vanish.
In Fig.~2, we show effects of
the vertex corrections on the D.C. conductivity by plotting
$\sigma_{VC}/\sigma$, where $\sigma_{VC}\;(\sigma)$ is the D.C. conductivity
as a function of $E_{H}/\gamma_{00}$
with (without) vertex corrections. We choose
$R_{2}=0.9$ and $R_{3}=0.8$.
The top panel is for the Born limit
while the bottom panel is for the unitary limit. As one can see, the vertex
corrections are significant particularly for the unitary limit. However,
in the presence of a sufficiently large field $(E_{H}\gg\gamma_{00})$,
the corrections become relatively unimportant. We found that the
vertex corrections to the
thermal conductivity are negligible in the presence of a magnetic field
as in the zero field case of Ref.\cite{durst}.

In conclusion, the WF law is violated in the
vortex state at zero temperature. The vertex corrections
to the conductivity become unimportant in the presence of
a sufficiently high field and may enhance the WF violation.
\begin{figure}[htb]
\begin{center}
\includegraphics[width=14pc,height=11pc]{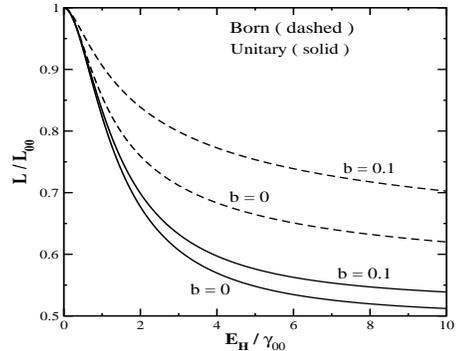}
\caption{Normalized Lorenz number
as a function of the magnetic energy normalized 
to the zero frequency and zero temperature
effective impurity scattering rate.
The parameter $b=0$ corresponds to
neglecting Andreev scattering from the random vortices.}
\label{fig1}
\end{center}
\end{figure}
\begin{figure}[htb]
\begin{center}
\includegraphics[width=14pc,height=14pc]{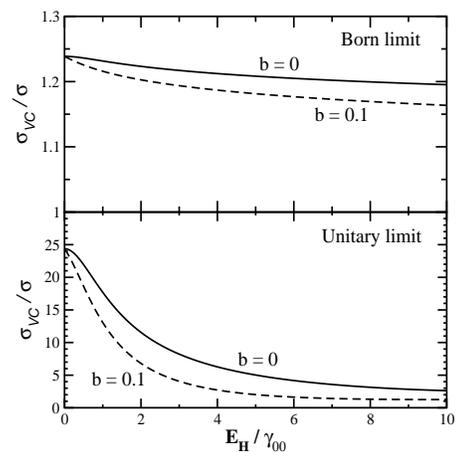}
\caption{Vertex corrections to the D.C. conductivity in the
Born (top panel) and unitary (bottom panel) limit. Notation as in Fig.~1.
}
\label{fig2}
\end{center}
\end{figure}

\end{document}